\documentclass[aps,prl,twocolumn,superscriptaddress,showpacs]{revtex4-1}

\usepackage[]{graphicx}
\usepackage[latin1]{inputenc}
\usepackage{color}
\usepackage{amssymb}
\usepackage{amsmath}

\begin{document}

\title{Direct Observation of Large Amplitude Spin Excitations Localized in a Spin-Transfer Nanocontact}
 
\author{D. Backes}
\email[]{dirk.backes@web.de}
\affiliation{Department of Physics, New York University, 4 Washington Place, New York, NY 10003, USA}
\author{F. Maci\`a}
\affiliation{Department of Physics, New York University, 4 Washington Place, New York, NY 10003, USA}
\affiliation{Grup de Magnetisme, Departament de F\'isica Fonamental, Universitat de Barcelona, Spain}
\author{S. Bonetti}
\affiliation{SLAC National Accelerator Laboratory, 2575 Sandhill Road, Menlo Park, CA 94025, USA}
\affiliation{Department of Physics, Stanford University, Stanford, CA 94305, USA}
\author{R. Kukreja}
\affiliation{SLAC National Accelerator Laboratory, 2575 Sandhill Road, Menlo Park, CA 94025, USA}
\affiliation{Department of Materials Science and Engineering, Stanford University, Stanford, CA 94305, USA}
\author{H. Ohldag}
\affiliation{Stanford Synchrotron Radiation Laboratory, 2575 Sandhill Road, Menlo Park, CA 94025, USA}
\author{A. D. Kent}
\affiliation{Department of Physics, New York University, 4 Washington Place, New York, NY 10003, USA}
\date{\today}

\begin{abstract}
We report the direct observation of large amplitude spin-excitations localized in a spin-transfer nanocontact using scanning transmission x-ray microscopy. Experiments were conducted using a nanocontact to an ultrathin ferromagnetic multilayer with perpendicular magnetic anisotropy. Element resolved x-ray magnetic circular dichroism images show an abrupt onset of spin excitations at a threshold current that are localized beneath the nanocontact, with average spin precession cone angles of $25^\circ$ at the contact center. The results strongly suggest that we have observed a localized magnetic soliton.
\end{abstract}

%
%

\maketitle

Spin-torque nano-oscillators (STNO) are nanometer scale contacts to thin magnetic layers that enable the generation of high current densities of spin-polarized electrons. Injection of spin polarized electrons into a ferromagnet leads to dynamic excitations of the magnetization associated with the generation of spin-waves. Electrical studies of STNO have indeed revealed such excitations at GHz frequencies \cite{Rippard2004,Silva2008}. In addition, an upper limit to the spatial extent of the spin excitation in layers with in-plane magnetic anisotropy has been measured in Brillouin light scattering experiments \cite{Demidov2010,Madami2011}. The relevant microscopic physical processes driving the dynamical behavior in STNO are of significant fundamental interest in this rapidly growing field, in particular considering its widespread potential for applications in the area of data storage and processing \cite{Macia2011}. However, a detailed microscopic understanding of spin transfer induced dynamics on the nanoscale is still elusive since it requires a direct quantitative magnetic characterization of the induced excitations at the relevant length and timescales. To address these open questions we investigated STNO with  perpendicular magnetic anisotropy using scanning transmission x-ray microscopy (STXM) and determined the spatial extent and magnitude of the spin excitations.

We have chosen an STNO with perpendicular magnetic anisotropy because it represents a well defined model system, where current direction, remanent field, and magnetic anisotropy field are aligned parallel. This is relevant because the applied field and ferromagnetic layer's magnetic anisotropy are predicted to determine the nature of the excited spin-wave modes \cite{Slavin2005,Hoefer2008,Hoefer2010,Bonettiprl2010}. In a magnetic layer with perpendicular magnetic anisotropy that is also magnetized perpendicular to the film plane, spin-waves excited by an electrical current have a frequency that is less than the lowest propagating spin-wave mode, which coincides approximately with the ferromagnetic resonance frequency and are therefore expected to be strictly localized in the contact region. It has been predicted, and inferred indirectly from electrical measurements, that these localized excitations are dissipative solitons, localized excitations that balance exchange and magnetic anisotropy forces \cite{Hoefer2010,Mohseni2013,DropletMacia2014}.

Here we report the direct observation of current induced spin excitations using synchrotron-based scanning transmission x-ray microscopy. X-ray magnetic circular dichroism (XMCD) is employed to detect changes in the average direction of the magnetization. XMCD is a common method to study magnetic properties in an element specific manner \cite{thole:92,chen:95}. It directly probes the spin polarization of the valence electronic states via x-ray induced excitation of core level electrons. Small changes of the magnetization of $10^{-4}$ or less can be recorded with a spatial resolution of about $30$ nm using  state of the art x-ray optics in combination with a synchrotron as a tunable, polarized and pulsed soft x-ray source \cite{Fischer:15}. In addition, a lock-in detection scheme enables time resolved studies ($\Delta t \simeq 50$ ps) and very high sensitivity \cite{SSRLSTXM, bonetti2015}. Finally, due to the ability of x-rays to penetrate a few micrometers of material we are able to study isolated, buried magnetic layers. Altogether these capabilities enable us to observe spin excitations in the magnetic region right {\em beneath} the nanocontact where the current is injected and in its direct vicinity. In our samples where the free magnetic layer exhibits perpendicular anisotropy we used this approach to observe changes in magnetization as a function of the current. Our STXM images reveal an abrupt onset of spin excitations that are highly localized---consistent with theoretical expectations for a localized magnetic soliton---with excitation amplitudes that correspond to average spin precession cone angles up to $25^\circ$ at the contact center.

Our STNO consist of Cu nanocontacts ($150$ nm in nominal diameter) to a CoNi multilayer with perpendicular anisotropy and an in-plane magnetized fixed layer (Permalloy), the same layer stack as those studied in Ref.~\citealp{DropletMacia2014}. The CoNi multilayer ($0.2$ nm Co$| 0.6$ nm Ni)$\times 6 | 0.2$ nm Co) and Permalloy are separated by 10 nm of Cu, which is sufficiently thick to completely decouple the layers magnetically. The layer stack was grown on $100$ nm thick SiN membranes, that are required as a transparent substrate for the soft x-ray transmission experiments. The membrane was coated with a $500$ nm thick Al layer on the backside to increase the thermal conductivity. Since the microscopy experiments were conducted in a vacuum environment the Al layer was crucial for the thermal stability of the device as we will show later. 

\begin{figure}
\includegraphics[width= \columnwidth]{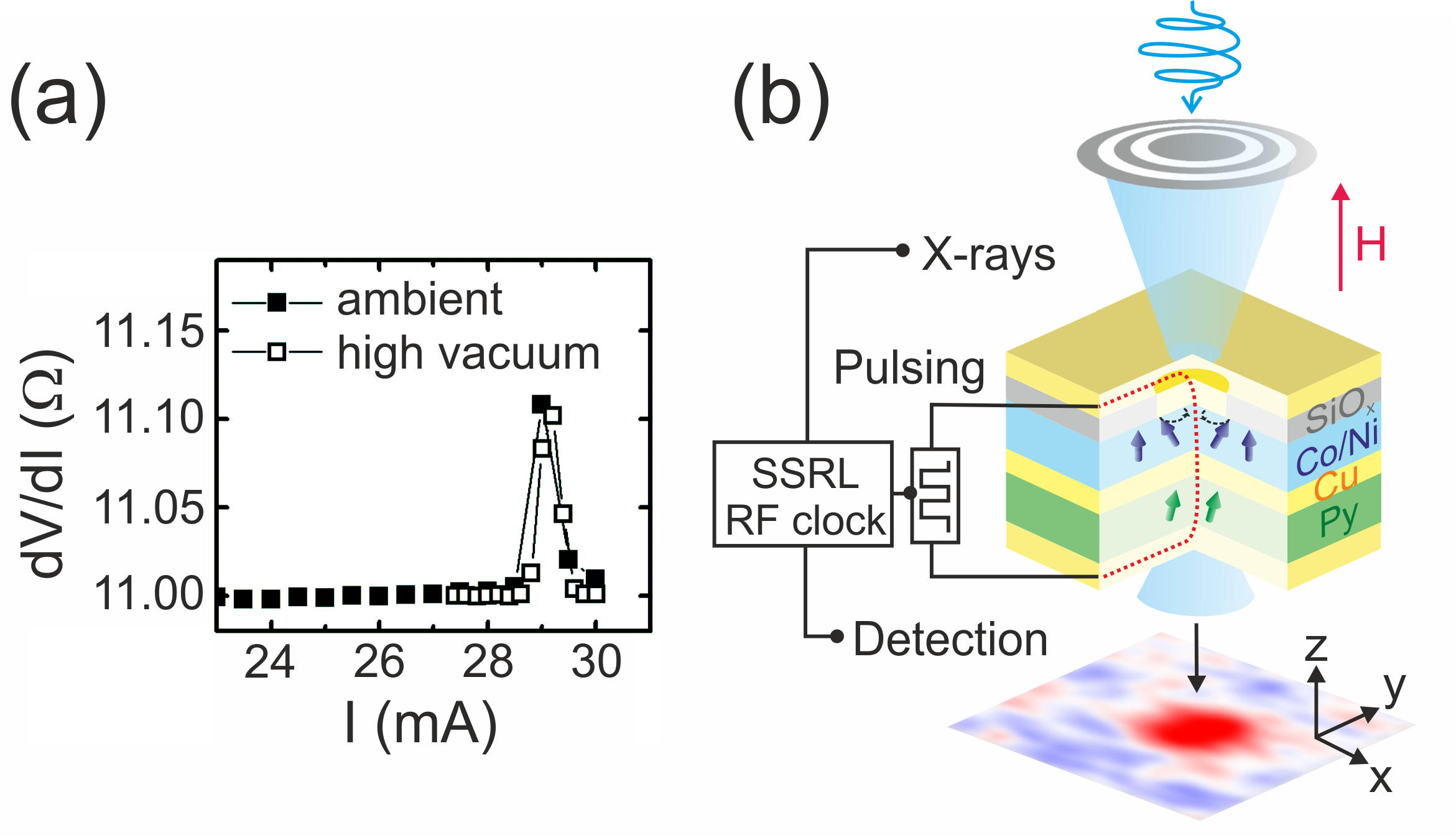}	
\caption{(a) STNO electrical characteristics: differential resistance $\mathrm{d}V/\mathrm{d}I$ versus current $I$. The peak at $29$ mA marks the threshold for current induced excitations. It occurs at the same current both in ambient conditions (filled squares) and high vacuum (open squares). (b) Schematic of the STXM instrument and the STNO sample. A Fresnel zone plate was used to focus the x-ray beam to a $35$ nm spot, which was scanned across the area around the nanocontact, indicated as the yellow region contacting the Co$|$Ni layer through the SiO$_2$ dielectric, to acquire an image. The x-ray detection was synchronized with the x-ray pulses from the synchrotron (RF clock) at $476.2$ MHz. }
\label{Fig1}
\end{figure}

We first characterized our samples ex-situ using ferromagnetic resonance spectroscopy, both directly after layer deposition and after STNO and membrane fabrication. The effective anisotropy field of the CoNi free layer was $\mu_\mathrm{0} H_\mathrm{eff}=\mu_\mathrm{0}(H_\mathrm{K}-M_\mathrm{s})=0.25$ T  \cite{Macia:jmmm2012}, with $\mu_\mathrm{0} H_\mathrm{K}=0.99$ T and $\mu_\mathrm{0} M_\mathrm{s}=0.74$ T, indicating a strong perpendicular magnetic anisotropy. Measurements before and after fabrication showed no change in the material properties. To further determine the current needed to excite magnetization dynamics we carried out electrical transport measurements. The electrical resistance between two magnetic layers across a non-magnetic layer depends on the relative orientation of their magnetizations due to the giant magnetoresistance effect. The onset of a magnetic excitation can therefore be detected by the presence of a peak in the differential $\mathrm{d}V/\mathrm{d}I$ \cite{DropletMacia2014}, since the average magnitude of a component of the magnetization changes. We then repeated the measurements in vacuum to corroborate that the Al layer serves as an effective thermal sink to counteract the reduced thermal conductivity in vacuum as designed. For this purpose the sample was mounted in the same configuration as used in the microscopy experiments. The two curves are shown in Fig. \ref{Fig1}(a). A pronounced peak appeared at a current of 29 mA, and no significant differences were observed between in-situ and ex-situ measurements.

To image the spin excitations we then used the STXM instrument at beamline 13-1 at the Stanford Synchrotron Radiation Lightsource (SSRL) \cite{SSRLSTXM, bonetti2015}. The incident photon energy was tuned to the Co L$_3$-edge (778.1 eV), to make use of the element specificity and only probe changes in the magnetization in the free layer, which is the only layer in the STNO that contains Co. The x-ray beam was aligned perpendicular to the sample surface as illustrated in Fig.~\ref{Fig1}b and a static magnetic field of $0.7$ T was applied perpendicular to the sample plane using a permanent magnet. As the absorption is proportional to the dot product of the magnetization $\bf{M}$ and the helicity $\bf{P}$ of the circularly polarized light \cite{thole:92}, the change of the perpendicular component of magnetization ($M_z$) can be determined in this geometry. The transmitted x-ray pulses were detected and amplified via an avalanche photodiode and registered using a software defined photon counting system \cite{Acremann:07}, that effectively acts as a lock-in amplifier operating at the x-ray pulse repetition rate of the synchrotron at 476.2 MHz. In addition, we modulated the applied current at 640 kHz, synchronized with  the frequency corresponding to the completion of one full electron orbit in the storage ring. We then compared the transmitted x-ray intensity for current on and off cycles, i.e. excitation on and off, for each image point. This double lock-in scheme allowed us to detect very small changes in the x-ray transmission ($<10^{-4}$), induced by the current by eliminating long term drifts and provided a reliable normalization scheme. 

Before we discuss the observed excitations we establish the effective magnetic thickness of the material by measuring the static XMCD effect of the Co layers. This will be important to evaluate the dynamic changes in magnetization in a quantitative manner later. We compared the transmission for positive (`$+$') and negative x-ray helicities (`$-$'), corresponding to parallel and antiparallel alignment between the magnetization and the polarization (i.e., $\bf{M}$ and $\bf{P}$). The ratio of the intensities is given by:
\begin{equation}
\ln(I_{+}^{*}/I_{-}^{*})=(\mu_+-\mu_-)t=\Delta \mu t
\end{equation}
where $I_{\pm}^{*}=I_{\pm}/I_{0,\pm}$ is the normalized beam intensity after transmission through the sample, $I_{0,\pm}$ the beam intensity, $\mu_{\pm}$ the spin-dependent absorption coefficient and $t$ the layer thickness. We obtained an XMCD contrast that corresponds to a Co thickness of $\sim 1.3$ nm, very close to the nominal Co thickness in the free layer of 1.4 nm.

\begin{figure}
\includegraphics[width= \columnwidth]{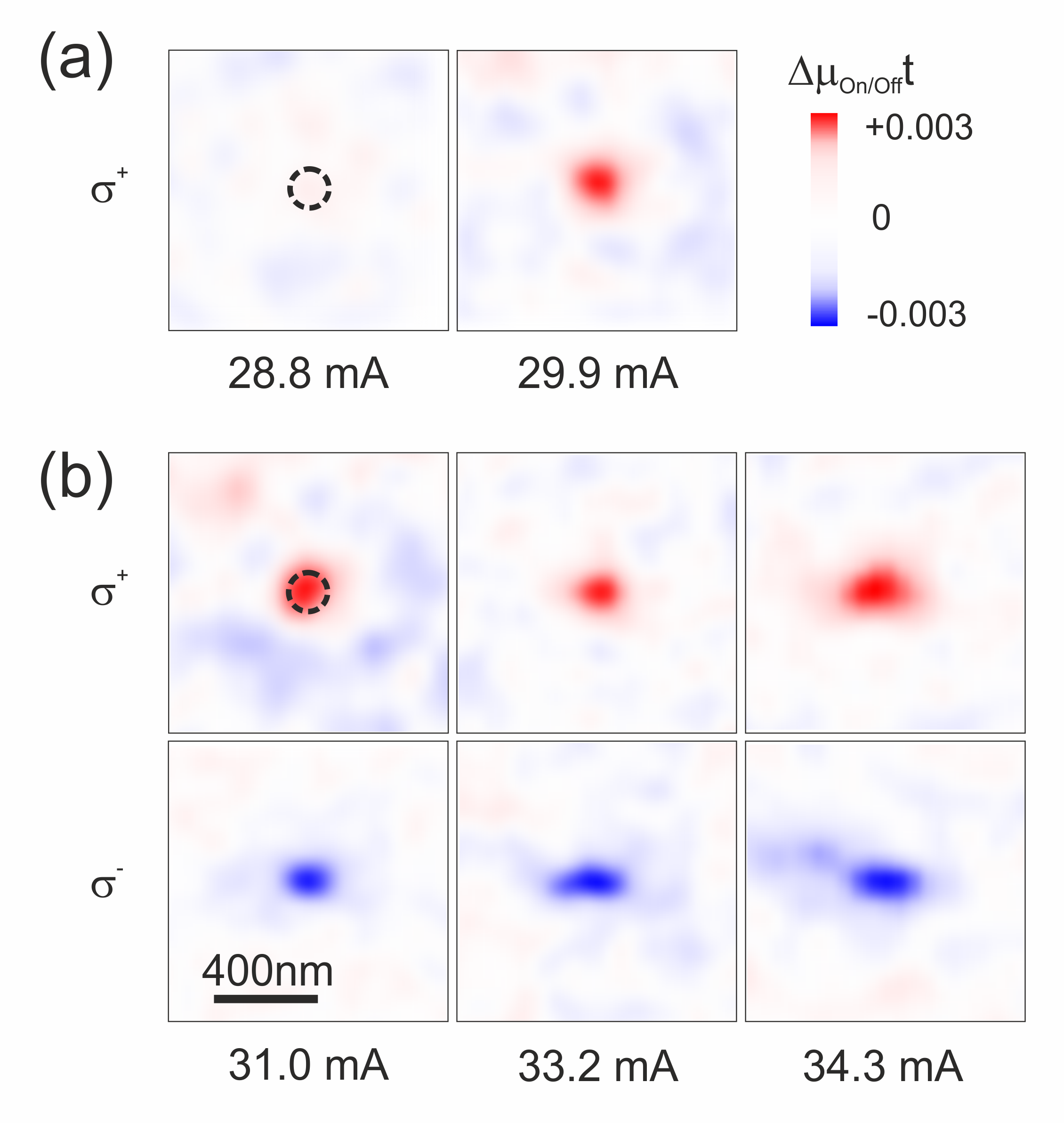}	
\caption{XMCD images of the nanocontact region at applied currents of (a) $28.8$ mA and $29.9$ mA, taken with positive helicity, and (b) $31.0$ mA, $33.2$ mA and $34.3$ mA, taken with both positive and negative x-ray helicities. The nanocontact is located in the center of each image, indicated with dashed circles in some of the images. The positive contrast in the nanocontact region for positive helicity and  negative contrast for negative helicity is consistent with a reduced magnetization component ($M_z$) in the contact region above the threshold current.}
\label{Fig2}
\end{figure}

We then recorded STXM images as a function of the applied current. Figs. \ \ref{Fig2}(a) and (b) show XMCD images with the nanocontact region outlined with a dashed line. For currents less than the $29$ mA we did not observe any XMCD contrast in our STXM images (see Fig.\ \ref{Fig2}(a)). However, at a current of $29.9$ mA we detected a pronounced excitation around the position of the nanocontact. This suggests that the observed feature appears abruptly at a current between $28.8$ mA and $29.9$ mA.  The fact that the observed contrast reverses its sign upon reversing the polarization is the signature of an XMCD effect, caused by a change in $M_z$. Fig.\ \ref{Fig2}(b) shows images at three different applied current values above the threshold current for both x-ray helicities; whereas `$+$' helicity shows an increase in the x-ray transmission (a red signal), `$-$' helicity, shows a decrease (a blue signal) in the x-ray transmission. This observation is consistent with a decrease of the average value of $M_z$. We observed magnetic dichroism in all images obtained with currents between $29.9$ mA to $34.3$ mA, the largest current we applied. Although the spatial extension of the observed excitations is mostly symmetric, some cases exhibit an elliptical deformation. We do not believe this represents a change in the vertical extend of the excitation. Considering that it takes 60 to 90 minutes to acquire a single image we attribute this deformation to small vertical drifts of the incident x-ray beam that cannot be compensated and lead to small changes in the vertical scale \cite{footnote1} 

\begin{figure}
\includegraphics[width= \columnwidth]{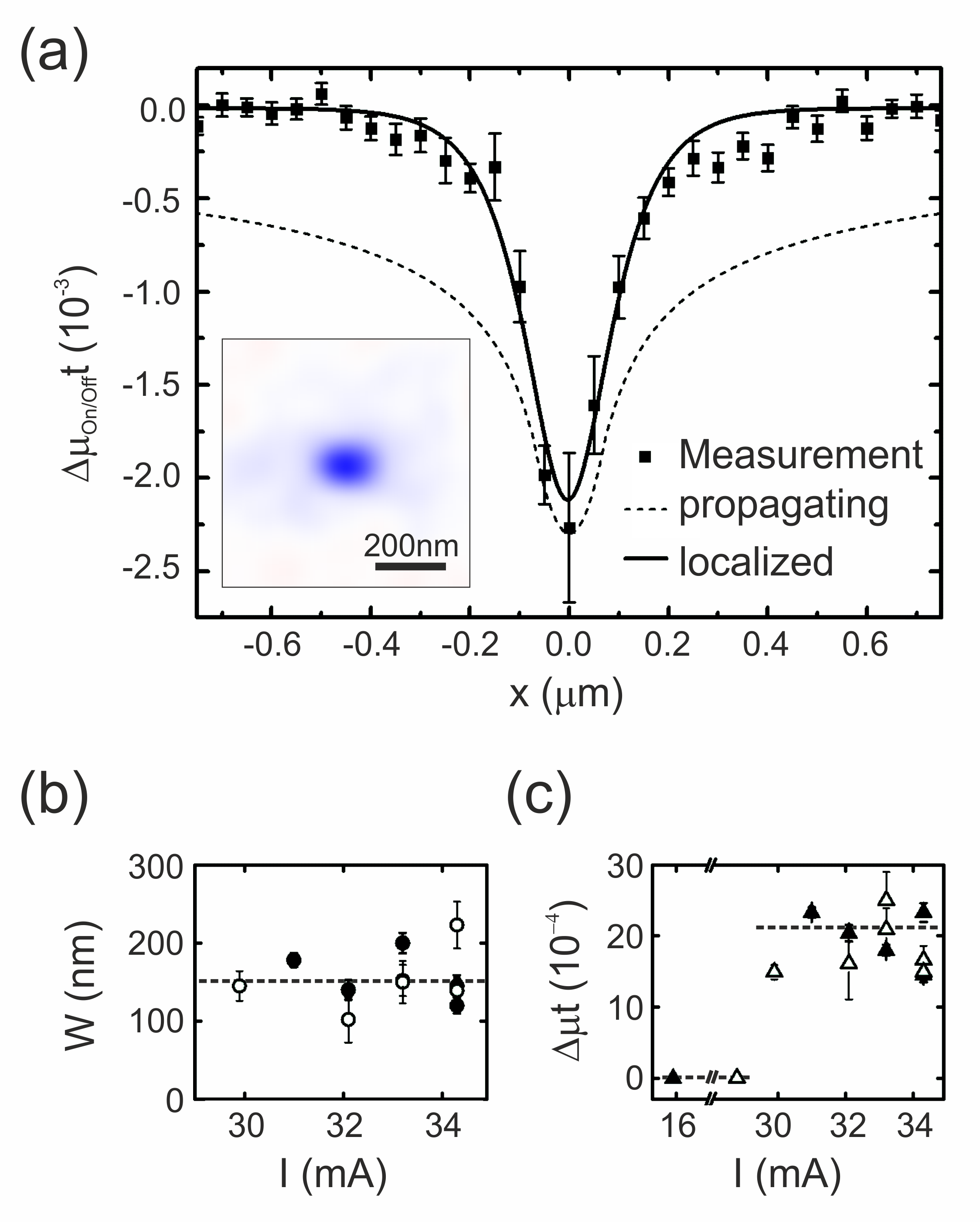}  
\caption{(a) Dynamic XMCD contrast (black squares) as a function of the distance $x$ from the nanocontact center for a current of $+31.0$ mA at negative x-ray helicity. The measurement is compared to a linear propagating mode (dashed line) and to a localized mode (straight line). (b) Width of the fitted localized modes (disks) and (c) amplitude (triangles) at different currents for negative (filled symbols) and positive (open symbols) x-ray helicity.}

\label{Fig3}
\end{figure}

We continue by quantitatively analyzing the image contrast by constructing one-dimensional profiles through the area of the  nanocontact. This is shown in Fig.\ \ref{Fig3}(a) for a current of $31.0$ mA in an image acquired with a negative x-ray helicity. Each point (black squares) represents an average over a half-circle at a certain distance to the right ($+x$) and left ($-x$) of the center of the nanocontact. We observe that the absorption signal decays rapidly outside of the nanocontact, having a full width at half maximum (FWHM) of $\simeq 175$ nm, just slightly larger than the nominal diameter of the contact ($150$ nm). It is instructive to compare the measured spin-wave excitation profiles to theoretical predictions. First we consider a propagating mode predicted by Slonczewski in a model that describes small amplitude excitations by linearizing the Landau-Lifschitz-Gilbert-Slonczewski (LLGS) torque equation \cite{Slonczewski1999}. This is shown as a dashed line in Fig.\ \ref{Fig3}(a). The envelope of the propagating mode clearly fails to describe the measured excitations, as it predicts a longer decay length and a larger excitation amplitude outside the contact region.  Also, the Slonczewski mode has only small amplitude excitations in the contact and our data show that the excitations have a large amplitude. Proposed corrections to the propagating modes that account for the nonlinearities \cite{Hoefer2005} show a similar (i.e. slow) decay and thus also do not fit our data.

Second, we plot the expected form of a soliton mode \cite{Kosevich1990}, a nonlinear, symmetric and localized mode (see the line in Fig.\ \ref{Fig3}(a)). This localized mode profile is a good fit to our data.  We used a hyperbolic secant as the profile for the soliton mode, derived from the LLGS equation for a perpendicular magnetized film  \cite{Kosevich1990}. The profile of this soliton mode is a good approximation to localized modes described as bullet modes for in-plane magnetized layers \cite{Slavin2005} and droplet soliton modes for layers with perpendicular magnetic anisotropy \cite{Hoefer2010}. We can also extract the mode amplitude that corresponds to the magnitude of the absorption at the contact center, and the mode width that characterizes the size of the localized excitation. This is shown in Fig.\ \ref{Fig3}(b) and (c) as a function of the applied current. The mode width fluctuates ($\simeq 175$ nm) with no particular trend as a function of the applied current. We believe that these variations are likely due to sample stage drift and our image processing: we measured images at the same currents more than once and obtained slightly different values, as shown in Fig.\ \ref{Fig3}(b).  By comparing the maximum amplitude of the excitations (see Fig.\ \ref{Fig3}(c)) to the absolute XMCD contrast (i.e., the contrast representing a 180$^{\circ}$ change of the magnetization), the precession cone angle $\theta_\mathrm{p}(r)$ can be determined from $\theta_\mathrm{p}=\arccos\left(1-\Delta\mu_\mathrm{on/off}(r)/\Delta\mu_{+/-}\right)$. Thus the amplitude of the peak in absorption indicates precession angles of about 25$^{\circ}$ at the center of the contact.

Several conclusions can be drawn from the analysis of the x-ray images. First, the observed excitation is localized at the nanocontact and exhibits almost circular symmetry. Second, its abrupt onset and large amplitude indicates that it is formed due to a nonlinear response of the magnetization to the applied spin-transfer torque. Third, the amplitude profile is consistent with that of a magnetic soliton as described by Kosevich {\em et al.} \cite{Kosevich1990}. More recently, Hoefer {\em et al.} presented a theory of dissipative droplet solitons in which they predicted an abrupt onset and large amplitude response for free layers with perpendicular magnetic anisotropy, which describes our experimental findings well \cite{Hoefer2010}. However, the authors also predicted a nearly complete reversal of the magnetization near the center of the contact that we have not observed. Our previous magnetoresistance measurements using the same type of samples at low temperature ($4.2$ K) and in the same range of applied fields ($\simeq 0.7$ T) indicate a nearly complete magnetization reversal in the contact region \cite{DropletMacia2014}. Measurements as a function of temperature up and above room temperature, showed a decreased response (step in contact $I-V$ characteristics and magnetoresistance) with increasing temperature \cite{Sergi}. This may indicate that thermal fluctuations play an important role in the magnetization dynamics. For instance the droplet may diffuse around the contact or even periodically annihilate and renucleate. The integration time in our STXM experiments is $500$ ms per point in the image and, as a result, changes in the envelope of the magnetization precession on shorter time scales than $500$ ms cannot be detected. 

In summary, we have observed large amplitude variations of the magnetic signal in an STNO with a CoNi multilayer with perpendicular magnetic anisotropy, having an equivalent Co thickness of only 1.4 nm, at a threshold current. Spatial images suggest that we have observed a localized magnetic soliton excited by spin-transfer torques. We determined the lower bound of the maximum angle of the excitation of $25^\circ$ and the upper bound of its spatial extent of $\sim 175$ nm, similar to the diameter of the STNO of $150$ nm. Our results also demonstrate the potential of STXM to resolve spin-wave excitations at nanometer length scales in specific magnetic layers in complex layer structures within nanostructured devices and provide a deeper understanding of the nature of current induced magnetic excitations.

\begin{acknowledgments}
We thank Mark Hoefer for his discussions of and comments on this manuscript. F.M. acknowledges support from the European Commission (Grant No. MC-IOF 253214), from Catalan Government COFUND-FP7, and from the Spanish Government (Grant No. MAT2011-23698.) S.B. gratefully acknowledges support from the Knut and Alice Wallenberg Foundation. This research was supported by NSF-DMR-1309202 and in part by ARO-MURI, Grant No. W911NF-08-1-0317. STNO were fabricated at the Center for Functional Nanomaterials, Brookhaven National Laboratory, which is supported by the U.S. Department of Energy, Office of Basic Energy Sciences, under Contract No. DE-AC02-98CH10886. Experiments at the Stanford Synchrotron Radiation Light Source, SLAC National Accelerator Laboratory are supported by the U.S. Department of Energy, Office of Basic Energy Sciences, under Contract No. DE-AC02-76SF00515.
\end{acknowledgments}


\begin{thebibliography}{22}
\bibitem{footnote1} We note that we observed a much smaller and less localized signal at negative applied currents, even for currents less than $29$ mA. However, the electrical measurements do not indicate that a negative current induces any magnetization dynamics. The x-ray data obtained with negative current is challenging to quantify, because of the weakness of the signal. We speculate that heating of the point contact due to the current could result in additional variations of the magnetization in the free layer, but due to the weakness of the signal it does not affect the interpretation of the data obtained with positive current.
\end{thebibliography}

\end{document}